\documentclass{PoS}

\newcommand{\beq}{\begin{equation}}
\newcommand{\eeq}{\end{equation}}
\newcommand{\bac}{\beq\begin{array}}
\newcommand{\eac}{\end{array}\eeq}
\newcommand{\ba}{\begin{array}}
\newcommand{\ea}{\end{array}}
\newcommand{\bea}{\begin{eqnarray}}
\newcommand{\eea}{\end{eqnarray}}

\title{The QCD Running Coupling and its Measurement}

\ShortTitle{Measurements of $\alpha_s$}

\author{\speaker{Guido Altarelli}\thanks{I am very grateful to George Zoupanos and the Organising Committee of the Corfu Summer Institute 2012 for their invitation and hospitality. This work has been partly supported by the Italian Ministero dell'Uni\-ver\-si\-t\`a e della Ricerca Scientifica, under the COFIN program (PRIN 2008), by the European Commission, under the networks ``LHCPHENONET'' and ``Invisibles'' }\\
       Dipartimento di Fisica `E.~Amaldi', Universit\`a di Roma Tre
\\
INFN, Sezione di Roma Tre, I-00146 Rome, Italy
\\
\vskip .1cm
and
\\
CERN, Department of Physics, Theory Division
\\
CH-1211 Geneva 23, Switzerland
\\

        E-mail: \email{guido.altarelli@cern.ch}}

\abstract{In this lecture, after recalling the basic definitions and facts about the running coupling in QCD, I present a critical discussion of the methods for measuring $\alpha_s$ and select those that appear to me as the most reliably precise}

\FullConference{RM3-TH/13-3; CERN-PH-TH/2013-059\\Proceedings of the Corfu Summer Institute 2012 "School and Workshops on Elementary Particle Physics and Gravity"\\
		September 8-27, 2012\\
		Corfu, Greece}

\begin{document}

\section{Introduction}
\label{sec:1}

The modern theory of strong interactions, QCD (Quantum Chromo-Dynamics) \cite{SM4,SM5,SM6}, has a simple structure but a very rich dynamical content (for books and reviews, see for example \cite{Book,collSM}). It gives rise to a
complex spectrum of hadrons, implies the striking properties of confinement and asymptotic freedom, is endowed with
an approximate chiral symmetry which is spontaneously broken, has a highly non trivial topological vacuum structure
(instantons, $U(1)_A$ symmetry breaking, strong CP violation (which is a problematic item in QCD possibly connected with new physics, like axions, ...), an intriguing phase transition diagram (colour
deconfinement, quark-gluon plasma, chiral symmetry restoration, colour superconductivity, ...). 

How do we get testable predictions from QCD? On the one hand there are non perturbative methods. The most important
at present is the technique of lattice simulations (for a recent review, see ref. \cite{kron}): it is based on first principles, it has produced very valuable
results on confinement, phase transitions, bound states, hadronic matrix elements and so on, and it is by now an
established basic tool. The main limitation is from computing power and therefore there is continuous progress and a lot
of good perspectives for the future. Another class of approaches is based on effective lagrangians which provide
simpler approximations than the full theory, but only valid in some definite domain of physical conditions.  Typically at energies below a given scale $L$ particles with mass larger than $L$ cannot be produced and thus only contribute short distance effects as virtual states in loops. Under suitable conditions one can write down a simplified effective lagrangian where the heavy fields have been eliminated (one says "integrated out"). Virtual heavy particle short distance effects are absorbed into the coefficients of the various operators in the effective Lagrangian. These coefficients are determined in a matching procedure, by requiring that the effective theory reproduces the matrix elements of the full theory up to power corrections. Important examples of effective theories are chiral
lagrangians, based on soft pion theorems \cite{chir} and valid for suitable processes at energies below 1~GeV (for a recent, concise review see ref. \cite{gas} and references therein); heavy
quark effective theories \cite{HQ}, obtained from expanding in inverse powers of the heavy quark mass,
mainly important for the study of  b and, to less accuracy, c decays (for reviews, see, for example, ref. \cite{neu}); soft-collinear effective theories (SCET) \cite{SCET}, valid for processes where quarks with energy much larger than their mass appear.  On the other hand, the perturbative approach, based on asymptotic freedom, still by far remains the main quantitative connection to experiment, due to its wide range of
applicability to all sorts of "hard" processes. 

The crucial ingredient of perturbative QCD is the "running" coupling $\alpha_s(Q)$, the scale dependent effective coupling that decreases logarithmically  with increasing values of the energy scale $Q$ and tends towards zero at infinity ("asymptotic freedom").  At large scales the detailed form of the running is prescribed by the QCD theory. The running coupling is a property of the theory and it enters in the perturbative expansion of any hard process. Some particularly simple and clear hard processes must be used to measure the running coupling at a suitable energy scale. Once the running coupling is measured it can be used to predict all observables related to a variety of hard processes. In particular QCD plays a crucial role in the physics at hadron colliders where many important processes, like jets at large $p_T$, heavy quark pair production, Higgs production and many more have rates that start at order $\alpha_s^2$. It is clear that the measurements of  $\alpha_s$  are very important to work out the relevant predictions of the theory. The error on the measured value of $\alpha_s$ is a main component of the theoretical error in all perturbative QCD predictions. In this article after recalling the basic definitions and facts about the running coupling we go to our main focus which is a critical discussion of the methods for precisely measuring $\alpha_s$ and the corresponding results. 

\section{The QCD running coupling}
\label{sec:2}

In the QCD lagrangian the only parameters with the dimension of energy are the quark masses (in units $\hbar=c=1)$. Thus at the classical level massless QCD is scale invariant. But the scale invariance of massless QCD does not remain true in the quantum theory. The scale symmetry of
the classical theory is unavoidably destroyed by the regularization and renormalization procedure which introduce a
dimensional parameter in the quantum version of the theory. When a symmetry of the classical theory is necessarily
destroyed by quantization, regularization and renormalization one talks of an "anomaly". So, in this sense, scale
invariance in massless QCD is anomalous.

While massless QCD is finally not scale invariant, the departures from scaling are asymptotically small, logarithmic and
computable. In massive QCD there are additional mass corrections suppressed by powers of m/E, where E is the energy scale (for processes that are non singular in the limit $m\rightarrow 0$). At the parton level (q and g) we can conceive to apply the asymptotic predictions
of massless QCD to processes and observables (we use the word "processes" for both) with the following properties ("hard
processes"). (a) All relevant energy variables must be large:
\beq
E_i~=~z_iQ,~~~~~~~~~Q>>m_j;~~~~~~~~~~z_i\rm{:scaling~variables~o(1)}
\label{hp}
\eeq
(b) There should be no infrared singularities (one talks of "infrared safe" processes). (c) The processes
concerned must be finite for $m\rightarrow 0$ (no mass singularities). To possibly satisfy these criteria processes must be
as "inclusive" as possible: one should include all final states with massless gluon emission and add all mass degenerate
final states (given that quarks are massless also $q-\bar q$ pairs can be massless if "collinear", that is moving together
in the same direction at the common speed of light).

In perturbative  QCD one computes inclusive rates for partons (the fields in the lagrangian, that
is, in QCD, quarks and gluons) and takes them as equal to rates for hadrons. Partons and hadrons are considered as two
equivalent sets of complete states. This is called "global duality" and it is rather safe in the rare instance of a totally
inclusive final state. It is less so for distributions, like distributions in the invariant mass M ("local duality") where
it can be reliable only if smeared over a sufficiently wide bin in M.

For the derivation of the running coupling the basic framework is the renormalization group formalism. In this section we denote with $\alpha$ either $\alpha_s$ in the case of QCD or its analogue in QED. For a sufficiently well defined Green function  $G$, the renormalization group equation (RGE) in massless QCD can be written as:
\beq
[\frac{\partial}{\partial\log{\mu^2}}~+~\beta(\alpha)\frac{\partial}{\partial \alpha}~+~
\gamma_G(\alpha)]G_{ren}~=~0\label{RGE2}\\
\eeq
where $\alpha$ is the renormalized coupling defined at the scale $\mu$ ($\mu$ cannot be zero because of infrared divergences) and
\beq
\beta(\alpha)~=~\frac{\partial\alpha}{\partial\log{\mu^2}}\label{beta}\\
\eeq
and
\beq
\gamma_G(\alpha)~=~\frac{\partial\log{Z_G }}{\partial\log{\mu^2}}\label{gamma}\\
\eeq
 Note that $\beta(\alpha)$ does not depend on which Green function $G$ we are considering; actually it is a property of the
theory and of the renormalization scheme adopted, while $\gamma_G(\alpha)$ also depends on $G$. Strictly speaking the RGE as written above is only valid in the Landau gauge ($\lambda=0$). In other gauges an additional term that takes the variation of the gauge fixing parameter $\lambda$ should also be included. We omit this term, for simplicity, as it is not relevant at the 1-loop level.

Assume that we want to apply the RGE to some hard process at a large scale $Q$, related to a Green function G that we can
always take as dimensionless (by multiplication by a suitable power of $Q$). Since the interesting dependence on $Q$ 
will be logarithmic we introduce the variable $t$ as :
\beq
t~=~\log{\frac{Q^2}{\mu^2}}\label{t}\\
\eeq
Then we can write $G_{ren}\equiv F(t,\alpha,x_i)$ where $x_i$ are scaling variables (we often omit to write them in the
following). In the naive scaling limit
$F$ should be independent of $t$, according to the classical intuition that massless QCD is scale invariant. To find the actual dependence on $t$, we want to solve the RGE
\beq
[-\frac{\partial}{\partial t}~+~\beta(\alpha)\frac{\partial}{\partial \alpha}~+~
\gamma_G(\alpha)]G_{ren}~=~0\label{RGE3}\\
\eeq
with a given boundary condition at $t=0$ (or $Q^2=\mu^2$): $F(0,\alpha)$. 
The solution of this general equation is given by:
\beq
F(t,\alpha)~=~F[0,\alpha(t)]\exp{\int_{\alpha}^{\alpha(t)}\frac{\gamma(\alpha')}{\beta(\alpha')}d\alpha'}\label{Fsol2}\\
\eeq
where the "running coupling" $\alpha(t)$ is defined by:
\beq
t~=~\int_{\alpha}^{\alpha(t)}\frac{1}{\beta(\alpha')}d\alpha'\label{run}\\
\eeq
Note that from this definition it follows that $\alpha(0)=\alpha$, so that the boundary condition is also satisfied.
To prove that $F(t,\alpha)$ in  eq. \ref{Fsol2} is indeed the solution, we first take derivatives with respect of $t$ and $\alpha$ (the two
independent variables) of both sides of eq.~(\ref{run}). By taking $d/dt$ we obtain
\beq
1~=~\frac{1}{\beta(\alpha(t)}\frac{\partial\alpha(t)}{\partial t}\label{ddt}\\
\eeq
We then take $d/d\alpha$ and obtain
\beq
0~=~-\frac{1}{\beta(\alpha)}~+~\frac{1}{\beta(\alpha(t)}\frac{\partial\alpha(t)}{\partial \alpha}\label{dda}\\
\eeq
These two relations make explicit the dependence of the running coupling on $t$ and $\alpha$:
\bea
\frac{\partial\alpha(t)}{\partial t}~=~\beta(\alpha(t))\label{runt}\\
\frac{\partial\alpha(t)}{\partial \alpha}~=~\frac{\beta(\alpha(t))}{\beta(\alpha)}\label{runa}\\
\nonumber
\eea
Using these two equations one immediately checks that $F(t,\alpha)$ is indeed the solution. 
In fact the sum of the two derivatives acting on the factor $F[0,\alpha(t)]$ vanishes and the exponential is by itself a
solution of the complete equation. Note that the boundary condition is also satisfied. The important point is the appearance of the running coupling as the quantity that determines the asymptotic departures from scaling. 

The
next step is to study the functional form of the running coupling. From eq.~(\ref{runt}) we see that the rate of change
with $t$ of the running coupling is determined by the $\beta$ function. In turn $\beta(\alpha)$ is determined by the $\mu$
dependence of the renormalized coupling through eq.~(\ref{beta}). Clearly there is no dependence on $\mu$ of the basic
3-gluon vertex in lowest order (order $e$). The dependence starts at 1-loop, that is at order $e^3$ (one extra gluon has
to be emitted and reabsorbed). Thus we obtain that in perturbation theory:
\beq
\frac{\partial e}{\partial \log{\mu^2}}~\propto~e^3\label{de}\\
\eeq
Recalling that $\alpha~=~e^2/4\pi$, we have:
\beq
\frac{\partial \alpha}{\partial \log{\mu^2}}~\propto~2e\frac{\partial e}{\partial \log{\mu^2}}~\propto~e^4
~\propto \alpha^2\label{da}\\
\eeq
Thus the behaviour of $\beta(\alpha)$ in perturbation theory is as follows:
\beq
\beta(\alpha)~=~\pm b\alpha^2[1~+~b'\alpha~+...]\label{betapert}\\
\eeq
Since the sign of the leading term is crucial in the following discussion, we stipulate that always $b>0$ and we
make the sign explicit in front. 

By direct calculation at 1-loop one finds:
\beq
\rm{QED:}~~~~~~~~\beta(\alpha)~\sim~+b\alpha^2~+.....~~~~~~~~~~~b~=~\sum_i\frac{N_CQ^2_i}{3\pi}
\label{beQED}\\
\eeq
where $N_C = 3$ for quarks and $N_C = 1$ for leptons and the sum runs over all fermions of charge $Q_ie$ that are coupled. Also, one finds:
\beq
\rm{QCD:}~~~~~~~~\beta(\alpha)~\sim~-b\alpha^2~+.....~~~~~~~~~~~b~=~\frac{11N_C-2n_f}{12\pi}\label{beQCD}\\
\eeq
where, as usual, $n_f$ is the number of coupled (see below) flavours of quarks (we assume here that $n_f~\le~16$ so that $b>0$ in QCD).
If
$\alpha(t)$ is small we can compute
$\beta(\alpha(t))$ in perturbation theory. The sign in front of $b$ then decides the slope of the coupling: $\alpha(t)$
increases with t (or
$Q^2$) if $\beta$ is positive at small $\alpha$ (QED), or $\alpha(t)$ decreases with t (or
$Q^2$) if $\beta$ is negative at small $\alpha$ (QCD). A theory like QCD where the running coupling vanishes
asymptotically at large $Q^2$ is called (ultraviolet) "asymptotically free". An important result that has been proven \cite{cogro} is
that in 4 spacetime dimensions all and only non-abelian gauge theories are asymptotically free.

Going back to eq.~(\ref{run}) we replace $\beta(\alpha)~\sim~\pm b\alpha^2$, do the integral and perform a simple algebra.
We find 
\beq
\rm{QED:}~~~~~~~~\alpha(t)~\sim~\frac{\alpha}{1-b\alpha t}\label{beQED1}\\
\eeq
and
\beq
\rm{QCD:}~~~~~~~~\alpha(t)~\sim~\frac{\alpha}{1+b\alpha t}\label{beQCD1}\\
\eeq
A slightly different form is often used in QCD. Defining $\Lambda_{QCD}$ by $1/\alpha~=~b\log{\mu^2/\Lambda_{QCD}^2}$, we can write:
\beq
\alpha(t)~\sim~\frac{1}{\frac{1}{\alpha}~+~bt}~=~\frac{1}{b\log{\frac{\mu^2}{\Lambda_{QCD}^2}}~+~b\log{\frac{Q^2}{\mu^2}}}
~=~\frac{1}{b\log{\frac{Q^2}{\Lambda_{QCD}^2}}}\label{alfaQCD}\\
\eeq
The parameter $\mu$ has been traded for the parameter $\Lambda_{QCD}$. We see that $\alpha(t)$ decreases logarithmically with $Q^2$ and that one can introduce a dimensional parameter
$\Lambda_{QCD}$ that replaces $\mu$. Often in the following we will simply write $\Lambda$ for $\Lambda_{QCD}$. Note that it is clear that $\Lambda$
depends on the particular definition of $\alpha$, not only through the defining scale $\mu$ but also on the renormalization
scheme adopted. It also depends on the number $n_f$ of coupled flavours through the parameter $b$, and in general through the
$\beta$ function. It is very important to note that QED and QCD
are theories with "decoupling":  up to the scale $Q$ only quarks with masses $m<<Q$
contribute to the running of $\alpha$. This is clearly very important, given that
all applications of perturbative QCD so far apply to energies below the top quark mass $m_t$. For the validity of the
decoupling theorem \cite{ApCa} it is necessary that the theory where all the heavy particle internal lines are eliminated is still
renormalizable and that the coupling constants do not vary with the mass. These requirements are true for heavy
quarks in QED and QCD, but are not true in the electroweak theory where the elimination of the top would violate $SU(2)$
symmetry (because the t and b left-handed quarks are in a doublet) and the quark couplings to the Higgs multiplet (hence to the
longitudinal gauge bosons) are proportional to the mass. In conclusion, in QED and QCD, quarks with $m>>Q$ do not contribute
to $n_f$ in the coefficients of the relevant $\beta$ function. The effects of heavy quarks are power suppressed and can be
taken separately into account. For example, in $e^+e^-$ annihilation for
$2m_c<Q<2m_b$ the relevant asymptotics is for $n_f=4$, while for $2m_b<Q<2m_t$ $n_f=5$. Going accross the $b$ threshold
the $\beta$ function coefficients change, so the $\alpha(t)$ slope changes. But $\alpha(t)$ is continuous, so that
$\Lambda$ changes so as to keep $\alpha(t)$ constant at the matching point at $Q\sim o(2m_b)$. The effect on $\Lambda$ is
large: approximately $\Lambda_5~\sim~0.65\Lambda_4$ where $\Lambda_{4,5}$ are for $n_f=4,5$. 

Note the presence of a pole in eqs.(\ref{beQED1},\ref{beQCD1}) at $\pm b\alpha t~=~1$, called the Landau pole, who realized
its existence in QED already in the '50's. For $\mu~\sim m_e$, in QED, the pole occurs beyond the Planck mass. In QCD the Landau
pole is located for negative $t$ or at $Q<\mu$ in the region of light hadron masses. Clearly the issue of the definition
and the behaviour of the physical coupling (which is always finite, when defined in terms of some physical process) in the region around the perturbative Landau pole is a problem that lies
outside the domain of perturbative QCD.
 
The non leading terms in the asymptotic behaviour of the running coupling can in principle be evaluated going back to
eq.~(\ref{betapert}) and computing $b'$ at 2-loops and so on. But in general the perturbative coefficients of
$\beta(\alpha)$ depend on the definition of the renormalized coupling $\alpha$ (the renormalization scheme), so one
wonders whether it is worthwhile to do a complicated calculation to get $b'$ if then it must be repeated for a
different definition or scheme. In this respect it is interesting to remark that actually one can easily prove that, due to the definition of $\beta(\alpha)$, both $b$ and $b'$ are independent
of the definition of $\alpha$, while higher order coefficients do depend on that.

In QCD ($N_C=3$)
one has obtained \cite{cas}:
\beq
b'~=~\frac{153-19n_f}{2\pi(33-2n_f)}\label{b'}\\
\eeq
By taking $b'$ into account one can write the expression of the running coupling at next to the leading order (NLO):
\beq
\alpha(Q^2)~=~\alpha_{LO}(Q^2)[1~-~b'\alpha_{LO}(Q^2)\log{\log{\frac{Q^2}{\Lambda^2}}}~+~...]\label{NLOa}\\
\eeq
where $\alpha_{LO}^{-1}~=~b\log{Q^2/\Lambda^2}$ is the LO result (actually at NLO the definition of $\Lambda$ is modified according to $b\log{\mu^2/\Lambda^2}=1/\alpha+b'\log{b\alpha}$).

At
present the universally adopted definition of $\alpha_s$ is in terms of dimensional regularization \cite{dimreg}, with the so-called Modified Minimal Subtraction
($\overline{MS} $) scheme \cite{MSbar} and a
value quoted for $\alpha_s$ is normally referring to this definition. Different measurements obtained at different energy scales are usually made comparable to each other by translating the result in terms of the value at the $Z$ mass: $\alpha_s(m_Z)$.

The third \cite{tar} and fourth \cite{ver} coefficients of the QCD $\beta$ function are also known in the $\overline{MS}$ prescription (recall that only the
first two coefficients are scheme independent). The calculation of the last term involved the evaluation of some 50,000 4-loop diagrams. Translated in numbers, for $n_f=5$ one obtains :
\beq
\beta(\alpha)~=~-0.610\alpha^2[1~+~1.261...\frac{\alpha}{\pi}~+~1.475...(\frac{\alpha}{\pi})^2~+~9.836...(\frac{\alpha}{\pi})^3...]\label{beta4}\\
\eeq 
It is interesting to remark that the expansion coefficients are of order 1 or 10 (only for the last one), so that the $\overline {MS}$ expansion looks
reasonably well behaved. 

Summarizing, massless classical QCD is scale invariant. But the procedure of
quantization, regularization and renormalization necessarily breaks scale invariance. In the quantum QCD theory there is a
scale of energy, $\Lambda$, which, from experiment, is of the order of a few hundred MeV, its precise value depending
on the definition, as we shall see in detail. Dimensionless quantities depend on the energy scale through the running
coupling which is a logarithmic function of
$Q^2/\Lambda^2$. In QCD the running coupling decreases logarithmically at large $Q^2$ (asymptotic freedom), while in QED
the coupling has the opposite behaviour.

\section{Measurements of $\alpha_s$}
\label{sec:3}

Very precise and reliable measurements of $\alpha_s(m_Z)$ are obtained from $e^+e^-$
colliders (in particular LEP), from deep inelastic scattering and from the hadron Colliders (Tevatron and LHC). The ''official''
compilation due to Bethke \cite{bethke} and included in the 2012 edition of the PDG \cite{pdg12} leads to the world average $\alpha_s(m_Z)=0.1184\pm0.0007$. A similar analysis with equivalent conclusions was recently presented in ref. \cite{pich}.
The agreement among so many different ways of measuring $\alpha_s$ is a strong quantitative test of QCD. However for some entries the stated error is taken directly from the original works and is not transparent enough as seen from outside (e.g. the lattice determination).
In my opinion one should select few theoretically simplest processes for measuring $\alpha_s$ and consider all other ways as
tests of the theory. Note that this is what is usually done in QED where $\alpha$ is measured from one single very precise and theoretically reliable observable (one possible calibration process is at present the electron g-2 \cite{ae1}).
The cleanest processes for measuring $\alpha_s$ are the totally inclusive ones (no hadronic corrections) with light cone dominance,
like Z decay, scaling violations in DIS and perhaps $\tau$ decay (but, for $\tau$, the energy scale is dangerously low). We will review these cleanest methods for measuring $\alpha_s$ in the following.

\subsection{$\alpha_s$ from $e^+e^-$ colliders}
\label{sec:3.1}

The totally inclusive processes for measuring $\alpha_s$ at $e^+e^-$ colliders are hadronic Z or $\tau$ decays ($R_{l,\tau}~=~\Gamma(Z,\tau\rightarrow hadrons)/\Gamma(Z,\tau\rightarrow
leptons)$, $\sigma_h$, $\sigma_l$, [$\sigma_F=12\pi\Gamma_l\Gamma_F/(m_Z^2\Gamma_Z^2)$ with F=h or l are the hadronic or leptonic cross sections at the $Z$ peak], and the total $Z$ width $\Gamma_Z=3\Gamma_l+\Gamma_h+\Gamma_{inv}$.
For each of this quantities, for example $R_l$, one can write a general expression of the form:
\beq
R_l~\sim~R^{EW}(1~+~\delta_{QCD}~+~\delta_{NP})~~\label{RR}\\
\eeq
where $R^{EW}$ is the electroweak-corrected Born approximation, $\delta_{QCD}$, $\delta_{NP}$ are the perturbative
(logarithmic) and non perturbative (power suppressed) QCD corrections. For a measurement of $\alpha_s$ (we always refer to the $\overline{MS}$ definition of $\alpha_s$) at the Z resonance peak  one can use all the information from $R_l$, $\Gamma_Z$ and 
$\sigma_{h,l}$. In the past the measurement from $R_l$ was preferred (by itself it leads to $\alpha_s(m_Z)=0.1226\pm0.0038$, a bit on the large side) but after LEP there is no reason for this preference. In all these quantities $\alpha_s$ enters through $\Gamma_h$, but the measurements of, say, $\Gamma_Z$, $R_l$ and $\sigma_l$ are really independent as they are affected by an entirely different systematics: $\Gamma_Z$ is extracted from the line shape, $R_l$ and $\sigma_l$ are measured at the peak but $R_l$ does not depend on the absolute luminosity while  $\sigma_l$ does. The most sensitive single quantity is $\sigma_l$. It gives $\alpha_s(m_Z)=0.1183\pm0.0030$. The combined value from the measurements at the Z (assuming the validity of the SM and the observed Higgs mass) is \cite{klu}:
\beq
\alpha_s(m_Z)=0.1187\pm0.0027\label{alZ}\\
\eeq
By adding all other electroweak precision electroweak tests (in particular $m_W$) one similarly finds \cite{ew}:
\beq
\alpha_s(m_Z)=0.1186\pm0.0026\label{alEW}\\
\eeq
These results have been obtained from the $\delta_{QCD}$ expansion up to and including the $c_3$ term of order $\alpha_s^3$. But by now the $c_4$ term (NNNLO!) has also been computed \cite{baik4} for inclusive hadronic $Z$ and $\tau$ decay. This remarkable calculation of about 20.000 diagrams, for the inclusive hadronic Z width, led to the result, for $n_f=5$ and $a_s=\alpha_s(m_Z)/\pi$:
\beq
\delta_{QCD}= [1~+~ a_s~+~0.76264~a_s^2  ~-~ 15.49~a_s^3~-~68.2~a_s^4~+~\dots]\\
\label{delc4}
\eeq
This result can be used to improve the value of  $\alpha_s(m_Z)$ from the EW fit given in eq. (\ref{alEW}) that becomes:
\beq
\alpha_s(m_Z)=0.1190\pm0.0026 \label{alEW1}\\
\eeq
Note that the error shown is dominated by the experimental errors. Ambiguities from higher perturbative orders \cite{ste}, from power corrections and also from uncertainties on the Bhabha luminometer (which affect $\sigma_{h,l}$) \cite{debo} are very small. In particular, having now fixed $m_H$ does not decrease the error significantly \cite{gfitter,gru}. The main source of error is the assumption of no new physics, for example in the $Zb\bar b$ vertex that could affect the $\Gamma_h$ prediction. 

We now consider the measurement of $\alpha_s(m_Z)$ from $\tau$ decay. $R_\tau$ has a number of advantages that, at least in
part, tend to compensate for the smallness of $m_\tau=1.777$~GeV. First, $R_\tau$ is maximally inclusive, more than
$R_{e^+e^-}(s)$, because one also integrates over all values of the invariant hadronic squared mass:
\beq
R_\tau=\frac{1}{\pi}\int_0^{m_\tau^2}\frac{ds}{m_\tau^2}(1-\frac{s}{m_\tau^2})^2 Im\Pi_\tau(s)\label{incl}\\
\eeq
As we have seen, the perturbative contribution is now known at NNNLO \cite{baik4}. Analyticity can be used to transform the integral into one on the circle at $|s|=m_\tau^2$:
\beq
R_\tau=\frac{1}{2\pi i}\oint_{|s|=m_\tau^2}\frac{ds}{m_\tau^2}(1-\frac{s}{m_\tau^2})^2 \Pi_\tau(s)\label{incl1}\\
\eeq  
Also, the factor $(1-\frac{s}{m_\tau^2})^2$ is important to kill the sensitivity the region $Re[s]=m_\tau^2$ where the
physical cut and the associated thresholds are located. Still the sensitivity to hadronic effects in the vicinity of the cut is a non negligible source of theoretical error that the formulation of duality violation models try to decrease. But the main feature that has attracted attention to $\tau$ decays for the measurement of  $\alpha_s(m_Z)$ is that even a rough determination of $\Lambda_{QCD}$ at a low scale $Q \sim m_\tau$ leads to a very precise prediction of  $\alpha_s$ at the scale $m_Z$, just because in $\log{Q/\Lambda_{QCD}}$ the value of  $\Lambda_{QCD}$ counts less and less as $Q$ increases. The absolute error on $\alpha_s$ shrinks by a factor of about one order of magnitude going from  $\alpha_s(m_\tau)$ to  $\alpha_s(m_Z)$. Still I find a little suspicious that to obtain a better measurement of  $\alpha_s(m_Z)$ you have to go down to lower and lower energy scales. And in fact, in general, in similar cases one finds that the decreased control of higher order perturbative and of non perturbative corrections makes the apparent advantage totally illusory. For $\alpha_s$ from $R_\tau$ the quoted amazing precision is obtained by taking for granted that corrections suppressed by $1/m_\tau^2$ are negligible. The argument is that in the massless theory, the light cone expansion is given by:
\beq
\delta_{NP}=\frac{ZERO}{m_\tau^2}~+~c_4\cdot \frac{<O_4>}{m_\tau^4}~+~c_6\cdot \frac{<O_6>}{m_\tau^6}~+\cdots\label{OPEtau}\\
\eeq
In fact there are no dim-2 Lorentz and gauge invariant operators. For example, $\sum_Ag^A_{\mu}g^{A\mu}$, where $g^A_\mu$ are the gluon fields, is not gauge invariant. In
the massive theory, the ZERO is replaced by the light quark mass-squared $m^2$. This is still negligible if $m$ is taken as a
lagrangian mass of a few MeV. If on the other hand the mass were taken to be the constituent mass 
of order $\Lambda_{QCD}$, this term would not be negligible at all and
would substantially affect the result (note that $\alpha_s(m_{\tau})/\pi\sim 0.1 \sim (0.6~{\rm GeV}/m_{\tau})^2$ and that $\Lambda_{QCD}$ for 3 flavours is large). The principle that coefficients in the operator expansion can be computed from the perturbative theory in terms of parton masses has never been really tested (due to ambiguities on the determination of condensates) and this particular case where a ZERO appears in the massless theory is unique in making the issue crucial. Many distinguished people believe the optimistic version. I am not convinced that
the gap is not filled up by ambiguities of $0(\Lambda_{QCD}^2/m_\tau^2)$ from $\delta_{pert}$: the $[ZERO/m_\tau^2]$ terms in eq. \ref{OPEtau} are vulnerable to possible ambiguities from the so-called ultraviolet renormalons  \cite{alnari}.
In fact one must keep in mind that the QED and QCD perturbative series, after renormalization, have all their
coefficients finite, but the expansion does not converge. Actually the perturbative series is not
even Borel summable (for reviews, see, for example refs. \cite{sixteen}). After the Borel resummation, for a given process one is left with a result which is ambiguous by terms
typically down by $\exp{-n/(b\alpha)}$, with $n$ an integer and b is the absolute value of the first $\beta$ function coefficient. In QED these
corrective terms are extremely small and not very important in practice. On the contrary in QCD $\alpha=\alpha_s(Q^2)\sim
1/(b\log{Q^2/\Lambda^2})$ and the ambiguous terms are of order $(1/Q^2)^n$, that is are power suppressed. It is interesting that, through this mechanism, the perturbative version of the theory is able to somehow take into account the power suppressed corrections. A sequence of diagrams with factorial growth at large order $n$ is made up by dressing  gluon propagators by any number of quark bubbles together with their gauge completions (renormalons).The problem of the precise relation between the ambiguities of the perturbative expansion and the higher twist corrections has been discussed in recent years \cite{sixteen}. 

There is a vast and sophisticated literature on $\alpha_s$ from $\tau$ decay. Unbelievably small errors on $\alpha_s(m_Z)$ are obtained in one or the other of several different procedures and assumptions that have been adopted to end up with a specified result. With time there has been an increasing awareness on the problem of controlling higher orders and non perturbative effects. In particular fixed order perturbation theory (FOPT) has been compared to resummation of leading beta function effects in the so called contour improved perturbation theory (CIPT). The results are sizably different in the two cases and there have been many arguments in the literature on which method is best. One important progress comes from the experimental measurement of moments of the $\tau$ decay mass distributions, defined by modifying the weight function in the integral in eq.(\ref{incl}). In principle one can measure  $\alpha_s$ from the sum rules obtained from different weight functions that emphasize different mass intervals and different operator dimensions in the light cone operator expansion. A thorough study of the dependence of the measured value of $\alpha_s$ on the choice of the weight function and in general of higher order and non perturbative corrections has appeared in ref.\cite{benbo} and I advise the interested reader to look at that paper and the references therein. It would be great to be able to fit both the value of $\alpha_s$ and of the coefficient of the $1/m_\tau^2$ term at the same time. Unfortunately a good precision on $\alpha_s$ is only obtained if that coefficient is fixed. 

We consider here the recent evaluations of  $\alpha_s$ from $\tau$ decay based on the NNNLO perturbative calculations \cite{baik4} and different procedures for the estimate of all sorts of corrections. From the papers given in refs. \cite{taulist} we obtain an average value and error that agrees with the Erler and Langacker values given in the PDG'12 \cite{pdg12}:
\beq
\alpha_s(m_{\tau})=0.3285\pm0.018\label{altautau}\\
\eeq
or
\beq
\alpha_s(m_Z)=0.1194\pm0.0021\label{altau}\\
\eeq
In any case, one can discuss
the error, but what is true and remarkable, is that the central value of $\alpha_s$ from $\tau$ decay, obtained at very small $Q^2$, is in
good agreement with all other precise determinations of $\alpha_s$ at more typical LEP values of $Q^2$. 

\subsection{$\alpha_s$ from Deep Inelastic Scattering}
\label{sec:3.2}

In principle DIS is expected to be an ideal laboratory for the determination of $\alpha_s$ but in practice the outcome is still to some extent open. QCD predicts the $Q^2$ dependence of $F(x,Q^2)$ at each fixed $x$, not the $x$ shape. But the $Q^2$ dependence is related to the
$x$ shape by the QCD evolution equations. For each x-bin the data allow to extract the slope of an approximately straight line
in $dlogF(x,Q^2)/dlogQ^2$: the log slope. The $Q^2$ span and the precision of the data are not much sensitive to the
curvature, for most $x$ values. A single value of $\Lambda_{QCD}$ must be fitted to reproduce the collection of the log
slopes. For the determination of $\alpha_s$ the scaling violations of non-singlet structure functions would be ideal,
because of the minimal impact of the choice of input parton densities. We can write the non-singlet evolution equations in
the form:
\beq
\frac{d}{dt}logF(x,t)~=~\frac{\alpha_s(t)}{2\pi}\int_x^1\frac{dy}{y}\frac{F(y,t)}{F(x,t)}P_{qq}(\frac{x}{y},\alpha_s(t))
\label{NSEE}\\
\eeq
where $P_{qq}$ is the splitting function. At present NLO and NNLO corrections are  known. It is clear from this form that, for example, the normalization error on the input
density drops away, and the dependence on the input is reduced to a minimum (indeed, only a single density appears here, while in
general there are quark and gluon densities). Unfortunately the data on non-singlet structure functions are not very
accurate. If we take the difference of data on protons and neutrons, $F_p-F_n$, experimental errors add up in the difference
and finally are large. The $F_{3\nu N}$ data are directly non-singlet but are not very precise. 
Another possibility is to neglect sea and glue in $F_2$ at sufficiently large $x$. But by only taking data at $x > x_0$ one decreases
the sample, introduces a dependence on $x_0$ and an
error from residual singlet terms. A recent fit to non singlet structure functions in electro- or muon-production extracted from proton and deuterium data, neglecting sea
and gluons at $x > 0.3$ (error to be evaluated) has led to the results \cite{blum}: 
\bea
\alpha_s(m_Z)&=&0.1148\pm0.0019 (exp) ~+~?~~~~~(NLO) \\
\alpha_s(m_Z)&=&0.1134\pm0.0020(exp) ~+~?~~~~~(NNLO) 
\label{alblu}
\eea
The central values are rather low and there is not much difference between NLO and NNLO. The question marks refer to the uncertainties from the residual singlet component at $x > 0.3$ and also to the fact that the old BCDMS data, whose systematics has been questioned, are very important at $x > 0.3$ and push the fit towards small values of $\alpha_s$.
 
When one measures $\alpha_s$ from scaling violations in $F_2$ measured with e or $\mu$ beams, the data are abundant, the statistical errors are small, the ambiguities from the treatment of heavy quarks and the effects of the longitudinal structure function $F_L$ can be controlled,  but there is an increased dependence on input parton densities and especially a strong correlation between the result on $\alpha_s$ and the adopted parametrization of the gluon density. In the following we restrict our attention to recent determinations of $\alpha_s$ from scaling violations at NNLO accuracy, as, for example, those in refs. \cite{ABM,J-Dr} that report the results, in the order:
\bea
\alpha_s(m_Z)&=&0.1134\pm0.0011(exp) ~+~?~~~~~\\
\alpha_s(m_Z)&=&0.1158\pm0.0035~~~~~
\label{recdet}
\eea 
In the first line my question mark refers to the issue of the $\alpha_s$-gluon correlation. In fact $\alpha_s$ tends to slide towards low values ($\alpha_s\sim 0.113-0.116$) if the gluon input problem is not fixed. Indeed, in the second line from ref. \cite{J-Dr}, the large error also includes an estimate of the ambiguity from the gluon density parametrization. One way to restrict the gluon density is to use the Tevatron high $p_T$ jet data to fix the gluon parton density at large $x$ that, via the momentum conservation sum rule, also constrain the small $x$ values of the same density. Of course in this way one has to go outside the pure domain of DIS. Also, the jet rates have been computed at NLO only. In a simultaneous fit of $\alpha_s$ and the parton densities from a set of data that, although dominated by DIS data, also contains Tevatron jets and Drell- Yan production, the result was \cite{MSTW}:
\beq
\alpha_s(m_Z)=0.1171\pm0.0014 ~+~?~~~~~\\
\label{mstw}
\eeq
The authors of ref. \cite{MSTW} attribute their larger value of $\alpha_s$ to a more flexible
parametrization of the gluon and the inclusion of Tevatron jet data that are important to fix the gluon at large $x$. An alternative way to cope with the gluon problem is to drastically suppress the gluon parametrization rigidity by adopting the neural network approach. With this method, in ref. \cite{NN}, from DIS data only, treated at NNLO accuracy, the following value was obtained:
\beq
\alpha_s(m_Z)=0.1166\pm0.0008(exp) \pm0.0009(th)~+~?~~~~\\
\label{nn1}
\eeq
where the stated theoretical error is that quoted by the authors within their framework, while the question mark has to do with possible additional systematics from the method adopted. Interestingly, in the same approach, by also including the Tevatron jets and the Drell-Yan data not much difference is found:
\beq
\alpha_s(m_Z)=0.1173\pm0.0007(exp) \pm0.0009(th)~+~?~~~~\\
\label{nn2}
\eeq
We see that when the gluon input problem is suitably addressed the fitted value of  $\alpha_s$ is increased.

As we have seen there is some spread of results, even among the most recent determinations based on NNLO splitting functions. We tend to favour determinations from the whole DIS set of data (i.e. beyond the pure non singlet case) and with attention paid to the gluon ambiguity problem (even if some non DIS data from Tevatron jets at NLO have to be included). A conservative proposal for the resulting value of $\alpha_s$ from DIS, that emerges from the above discussion is something like:  
\beq
\alpha_s(m_Z)=0.1165\pm0.0020~~~~~~\\
\label{myave}
\eeq
The central value is below those obtained from $Z$ and $\tau$ decays but perfectly compatible with those results.

\subsection{Other $\alpha_s(m_Z)$ Measurements as QCD Tests}
\label{sec:3.3}

There are a number of other determinations of $\alpha_s$ that are important because they arise from qualitatively different observables and methods. All together they provide an impressive set of QCD tests. Here I will give a few examples of these interesting measurements.

A classic set of measurements is from a number of infrared safe observables related to event rates and jet shapes in $e^+e^-$ annihilation. One important feature of these measurements is that they can be repeated at different energies in the same detector, like the JADE detector in the energy range of PETRA or the LEP detectors from LEP1 to LEP2 energies. As a result one obtains a striking direct confirmation of the running of the coupling according to the renormalization group prediction. The perturbative part is known at NNLO \cite{gher} and resummations of leading logs arising from the vicinity of cuts and/or boundaries have been performed in many cases using effective field theory methods. The main problem of these measurements is the possible large impact of non perturbative hadronization effects on the result and therefore on the theoretical error. 
According to ref.\cite{bethke} a summary result that takes into account the central values and the spread from the JADE measurements, in the range 14 to 46 GeV, at PETRA is given by: $\alpha_s(m_Z)=0.1172\pm 0.0051$, while from the ALEPH data at LEP, in the range 90 to 206 GeV, the reported value \cite{diss} is $\alpha_s(m_Z)=0.1224\pm 0.0039$.
It is amazing to note that among the related works there are a couple of papers by Abbate et al \cite{abb1,abb2} where an extremely sophisticated formalism is developed for the thrust distribution, based on NNLO perturbation theory with resummations at NNNLL plus a data/theory based estimate of non perturbative corrections. The final quoted results are unbelievably precise: $\alpha_s(m_Z)=0.1135\pm 0.0011$ from the tail of the Thrust distribution \cite{abb1} and  $\alpha_s(m_Z)=0.1140\pm 0.0015$ from the first moment of the Thrust distribution \cite{abb2} (note the low central value). I think that this is a good example of an underestimated error which is obtained within a given machinery without considering the limits of the method itself. 
Another allegedly very precise determination of $\alpha_s(m_Z)$ is obtained from lattice QCD by several groups \cite{lat} with different methods and compatible results. A value that summarizes these different results is \cite{pdg12} $\alpha_s(m_Z)=0.1185\pm 0.0007$. With all due respect to lattice people I think this small error is totally umplausible. 

\section{Conclusion: My Recommended Value of $\alpha_s(m_Z)$}
\label{sec:4}

According to my proposal to calibrate $\alpha_s(m_Z)$ from the theoretically cleanest and most transparent methods, identified as the totally inclusive, light cone operator expansion dominated processes, I collect here my understanding of the results:
from $Z$ decays and EW precision tests, eq.(\ref{alEW}):
\beq
\alpha_s(m_Z)=0.1190\pm0.0026; \label{alEW1pr}\\
\eeq
from scaling violations in DIS, eq.(\ref{myave}):
\beq
\alpha_s(m_Z)=0.1165\pm0.0020;~~~~~~\\
\label{myavepr}
\eeq
from $R_\tau$, eq.(\ref{altau}):
\beq
\alpha_s(m_Z)=0.1194\pm0.0021.\label{altaupr}\\
\eeq

If one wants to be on the safest side one can take the average of $Z$ decay and DIS:
\beq
\alpha_s(m_Z)=0.1174\pm0.0016.~~~~~~\\
\label{finave1}
\eeq
This is my recommended value. If one adds to the average the relatively conservative $R_\tau$ value and error given above in eq. \ref{altau}, that takes into account the dangerous low energy scale of the process, one obtains:
\beq
\alpha_s(m_Z)=0.1184\pm0.0011.~~~~~~\\
\label{finave2}
\eeq
Note that this is essentially coincident with the "official" average with a moderate increase of the error. 
Thus we see that a sufficiently precise measure of $\alpha_s(m_Z)$ can be obtained, eqs. (\ref{finave1},\ref{finave2}), by only using the simplest processes where the control of theoretical errors is maximal. One is left free to judge whether a further restriction of theoretical errors is really on solid ground.

The value of $\Lambda$ (for $n_f=5$) which corresponds to eq.~(\ref{finave1}) is:
\beq
\Lambda_5=202\pm 18~{\rm MeV}\label{lambda}\\
\eeq
while the value from eq.~(\ref{finave2}) is:
\beq
\Lambda_5=213\pm 13~{\rm MeV}\label{lambda2}\\
\eeq

In conclusion, we see that a reasonably precise measure of $\alpha_s(m_Z)$ can be obtained, eqs. (\ref{finave1},\ref{finave2}), by only using the simplest processes where the control of theoretical errors is maximal. One is left free to judge whether a further decrease of the error is really worthwhile at the price of a loss in transparency and rigour.


\begin{thebibliography}{99}

\bibitem{SM4} M. Gell-Mann, Acta Phys. Austriaca Suppl. IX (1972) 733; H. Fritzsch and M. Gell-Mann, Proc. XVI Int. Conf. on High Energy Physics, Chicago-Batavia, 1972; H. Fritzsch, M. Gell-Mann and H. Leutwyler, Phys. Lett. B {\bf 47}, 365 (1973)\\[-18pt]
\bibitem{SM5} D. Gross and F. Wilczek, Phys. Rev. Lett. {\bf 30}, 1343 (1973), Phys. Rev. D {\bf 8}, 3633 (1973); H.D. Politzer,  Phys. Rev. Lett. {\bf 30}, 1346 (1973)\\[-18pt]
\bibitem{SM6} S. Weinberg,  Phys. Rev. Lett. {\bf 31}, 494 (1973)\\[-18pt]
\bibitem{Book} 
Y.L. Dokshitzer and V.A. Khoze, {\em Basics of Perturbative QCD}, Ed. Frontieres (1991);  T. Muta, 
{\em Foundation of Quantum Chromodynamics: an Introduction to Perturbative Methods in Gauge Theories}, World Sci., 3rd ed. (2009); F. J. Yndurain, {\em The Theory of Quark and Gluon Interactions}, Springer, 4th ed., (2006). G. Altarelli, {\em The Development of Perturbative QCD}, World Scientific, (1994); W. Greiner, S. Schramm and E. Stein, {\em Quantum Chromodynamics}, Springer, 3rd ed. (2007); {\em Handbuch of QCD}, ed. by M. Shifman, Vol.1-4, World Sci. (2001); R. K. Ellis, W. J. Stirling, and B. R. Webber,  {\em QCD and Collider Physics}, Cambridge Monographs, (2003);  G. Dissertori, I. Knowles and M. Schmelling, {\em Quantum Chromodynamics: High Energy Experiments and Theory}, Oxford Univ. Press, 2003: S. Narison,{\em QCD as a Theory of Hadrons: From Partons to Confinement}, Cambridge Monographs, (2007); J. C. Collins, 
{\em Foundations of Perturbative QCD }, Cambridge Monographs, (2011)\\[-18pt]
\bibitem{collSM} G. Altarelli, ArXiv:1303.2842 \\[-18pt]
\bibitem{kron} A. S. Kronfeld, ArXiv:1203.1204 \\[-18pt]
\bibitem{chir}S. Weinberg, Physica A 96 (1979) 327, J. Gasser and H. Leutwyler,  Annals Phys. 158 (1984) 142; Nucl.Phys.B250(1985)465;J. Bijnens, G. Ecker and J. Gasser, hep-ph/9411232, in The Second DA$\Phi$NE Physics Handbook (Frascati, 1995); J. Bijnens and U. Meissner, hep-ph/9901381; G. Ecker, hep-ph/9805500, hep-ph/0011026; H. Leutwyler, hep-ph/9609465, hep-ph/0008124; A. Pich, hep-ph/9806303; E. de Rafael, hep-ph/9502254; L. Maiani, G. Pancheri and N. Paver, The Second DA$\Phi$NE Physics Handbook (Frascati, 1995); G. Ecker, Prog. Part. Nucl. Phys. 35 (1995) 1; G. Colangelo1 and G. Isidori,  ArXiv:hep-ph/0101264\\[-18pt]
\bibitem{gas} J. Gasser, Nucl.Phys.Proc.Suppl.86:257 (2000), ArXiv:hep-ph/9912548\\[-18pt]
\bibitem{HQ} E. Eichten and B. Hill, Phys. Lett. B 234, 511 (1990); H. Georgi, Phys. Lett. B 240, 447 (1990); B. Grinstein, Nucl. Phys. B 339, 253 (1990); T. Mannel, W. Roberts, and Z. Ryzak, Nucl. Phys. B 368, 204 (1992)\\[-18pt] 
\bibitem{neu} M. Neubert, Phys. Rept. 245, 259 (1994) ArXiv:hep-ph/9306320; A.V. Manohar and M.B. Wise, Camb. Monogr. Part. Phys. Nucl. Phys. Cosmol. 10, 1 (2000)\\[-18pt] 
\bibitem{SCET} C. W. Bauer, S. Fleming, and M.E. Luke, Phys. Rev. D 63, 014006 (2000),
ArXiv:hep-ph/0005275; C.W. Bauer et al., Phys. Rev. D 63, 114020 (2001) ArXiv:hep-ph/0011336]; C.W. Bauer and I.W. Stewart, Phys. Lett. B 516, 134 (2001) ArXiv:hep-ph/0107001; C.W. Bauer, D. Pirjol, and I.W. Stewart, Phys. Rev. D 65, 054022 (2002), ArXiv:hep-ph/0109045\\[-18pt]
\bibitem{cogro} S. Coleman and D. Gross, Phys. Rev. Lett. 31 (1973) 851\\[-18pt] 
\bibitem{ApCa} T. Applequist and J. Carazzone, Phys. Rev. D11 (1975)2856\\[-18pt]
\bibitem{dimreg} G. 't Hooft and M. Veltman, Nucl. Phys. B44 (1972) 189; C.G. Bollini and J.J. Giambiagi, Nuovo Cimento 12B (1972) 20;
J.F. Ashmore, Nuovo Cimento Lett. 4 (1972) 289; G.M. Cicuta and E. Montaldi, Nuovo Cimento Lett 4 (1972) 329\\[-18pt]
\bibitem{MSbar} G. 't Hooft, Nucl. Phys. B61 (1973) 455; W. A. Bardeen, A. J. Buras, D. W. Duke and T. Muta, Phys. Rev. D18(1978) 3998\\[-18pt]
\bibitem{cas} W. Caswell, Phys. Rev. Letters 33 (1974) 244; D. R. T. Jones, Nucl. Phys. B75 (1974) 531\\[-18pt]
\bibitem{tar} O.V. Tarasov, A.A. Vladimirov and A. Yu. Zharkov, Phys. Lett. 93B(1980)429\\[-18pt]
\bibitem{ver} T. van Ritbergen, J.A.M. Vermaseren and S.A. Larin, Phys.Lett. B400(1997)379; see also M. Czakon, Nucl. Phys. B710 (2005) 485\\[-18pt]
\bibitem{pdg12} Particle Data Group, J. Beringer et al, Phys. Rev. D{\bf 86}(2012) 010001\\[-18pt]
\bibitem{bethke} S. Bethke, ArXiv:0908.1135\\[-18pt]
\bibitem{pich} A. Pich, ArXiv:1303.2262\\[-18pt]
\bibitem{ae1} D. Hanneke, S. Fogwell, and G. Gabrielse, Phys. Rev. Lett. 100, 120801 (2008);
D. Hanneke, S. Fogwell Hoogerheide, and G. Gabrielse, Phys. Rev. A 83, 052122 (2011)\\[-18pt]
\bibitem{klu} S. Kluth, Rept. Prog. Phys. 69 (2006) 1771\\[-18pt]
\bibitem{ew} The LEP Electroweak Working Group, http://lepewwg.web.cern.ch/LEPEWWG/\\[-18pt]
\bibitem{baik4}P.A. Baikov, K.G. Chetyrkin and J.H. Kuhn, Phys. Rev. Lett. 101 (2008) 012002, ArXiv:0801.1821; Phys. Rev. Lett. 104 (2010) 132004, ArXiv:1001.3606P; A. Baikov, K. G. Chetyrkin, J. H. Kuhn and J. Rittinger, ArXiv:1210.3594\\[-18pt]
\bibitem{ste} H. Stenzel, JHEP 0507:0132,2005\\[-18pt]
\bibitem{debo} W. de Boer and C. Sander, Phys.Lett.B585:276,2004\\[-18pt]
\bibitem{gfitter} M. Baak et al, Gfitter group, ArXiv:1209.2716\\[-18pt]
\bibitem{gru} M. Grunewald, for the LEP EW Group, private communication\\[-18pt]
\bibitem{alnari} G Altarelli, P. Nason and G. Ridolfi, Z.Phys.C68:257,1995\\[-18pt]
\bibitem{sixteen}G. Altarelli, Proceedings of the E. Majorana Summer School, Erice, 1995, Plenum Press, ed. by A. Zichichi; M. Beneke and V. M. Braun, in {\em Handbuch of QCD}. ed. by M. Shifman, Vol.3, pag. 1719, World Sci. (2001)\\[-18pt]
\bibitem{benbo} M. Beneke, D. Boito and M. Jamin,  ArXiv:1210.8038\\[-18pt]
\bibitem{taulist}  M. Beneke and M. Jamin, JHEP 0809, 044 (2008),
ArXiv:0806.3156; M. Davier et al., Eur. Phys. J. C56, 305 (2008), ArXiv:0803.0979;  K. Maltman and T. Yavin, Phys. Rev. D78, 094020 (2008), ArXiv:0807.0650; S. Narison, Phys. Lett. B673, 30 (2009), ArXiv:0901.3823;
I. Caprini and J. Fischer, Eur. Phys. J. C64, 35 (2009); ArXiv:0906.5211; S. Menke, ArXiv:0904.1796; A. Pich, ;ArXiv:1107.1123; B.A. Magradze, ArXiv:1112.5958; G. Abbas et al., ArXiv:1202.2672; D. Boito et al., Phys. Rev. D84, 113006 (2011), ArXiv:1110.1127; D. Boito et al., ArXiv:1203.3146\\[-18pt]
\bibitem{blum} J. Blumlein, H. Bottcher and A. Guffanti, Nucl.Phys.B774:182,2007, 
ArXiv:0607200 [hep-ph]\\[-18pt]
\bibitem{ABM} S. Alekhin, J. Blumlein and S. Moch, ArXiv:1202.2281 [hep-ph]\\[-18pt] 
\bibitem{J-Dr} P. Jimenez-Delgado and E. Reya, Phys. Rev. D 79 (2009) 074023, ArXiv:0810.4274\\[-18pt] 
\bibitem{MSTW} A. D. Martin, W. J. Stirling, R. S. Thorne and G. Watt, Eur. Phys. J. C 64 (2009) 653, ArXiv:0905.3531\\[-18pt]
\bibitem{NN} R. D. Ball, et al. Phys. Lett. B 707 (2012) 66, ArXiv:1110.2483\\[-18pt]
\bibitem{gher} A. Gehrmann-De Ridder, T. Gehrmann, E.W.N. Glover and G. Heinrich, Phys. Rev. Lett. 99 (2007) 132002, ArXiv:0707.1285;
JHEP 0711 (2007) 058, ArXiv:0710.0346; Phys. Rev. Lett. 100 (2008) 172001, ArXiv:0802.0813, JHEP 0712 (2007) 094 [ArXiv:0711.4711\\[-18pt]
\bibitem{diss} G. Dissertori et al,  JHEP 0908:036,2009, ArXiv:0906.3436\\[-18pt]
\bibitem{abb1} R. Abbate, M. Fickinger, A. H. Hoang, V. Mateu and I. W. Stewart, Phys.Rev. D83 (2011) 074021, ArXiv:1006.3080\\[-18pt] 
\bibitem{abb2} R. Abbate, M. Fickinger, A. H. Hoang, V. Mateu and I. W. Stewart, Phys.Rev. D86 (2012) 094002, ArXiv:1204.5746\\[-18pt]
\bibitem{lat} C. McNeile et al., [HPQCD Collab.], Phys. Rev. D82, 034512 (2010)
ArXiv:1004.4285; C.T.H. Davies et al., [HPQCD Collab., UKQCD Collab., and MILC Collab.], Phys.
Rev. Lett. 92, 022001 (2004), ArXiv:0304004; Q. Mason et al, [HPQCD and UKQCD Collaborations], Phys. Rev. Lett. 95 (2005)052002, hep-lat/0503005; K. Maltman, et al., Phys. Rev. D78, 114504 (2008), ArXiv:0807.2020; S. Aoki et al., [PACS-CS Collab.], JHEP 0910, 053 (2009), ArXiv:0906.3906; E. Shintani et al., [JLQCD Collab.], Phys. Rev. D82, 074505 (2010), ArXiv:1002.0371; B. Blossier et al., [ETM Collab.], ArXiv:1201.5770\\[-18pt]




\end{thebibliography}
\end{document}